\documentclass[twocolumn,showpacs,preprintnumbers,amsmath,amssymb]{revtex4}
\usepackage{graphicx}% Include figure files
\usepackage{dcolumn}% Align table columns on decimal point
\usepackage{bm}% bold math

\begin{document}

\title{Bosonic characters of atomic Cooper pairs across resonance}

% Force line breaks with \\

\author{Y. H. Pong and C. K. Law}
\affiliation{Department of Physics and Institute of Theoretical
Physics, The Chinese university of Hong Kong, Shatin, Hong Kong SAR,
China}

\date{\today}
\begin{abstract}
We study the two-particle wave function of paired atoms in a Fermi
gas with tunable interaction strengths controlled by Feshbach
resonance. The Cooper pair wave function is examined for its bosonic
characters, which is quantified by the correction of Bose
enhancement factor associated with the creation and annihilation
composite particle operators. An example is given for a
three-dimensional uniform gas. Two definitions of Cooper pair wave
function are examined. One of which is chosen to reflect the
off-diagonal long range order (ODLRO). Another one corresponds to a
pair projection of a BCS state. On the side with negative scattering
length, we found that paired atoms described by ODLRO are more
bosonic than the pair projected definition. It is also found that at
$(k_F a)^{-1} \ge 1$, both definitions give similar results, where
more than 90\% of the atoms occupy the corresponding molecular
condensates.
\end{abstract}


\maketitle

\section{Introduction}
Recent advancement in the control of Feshbach molecules has given
rise to many new experimental observations in the world of ultracold
atomic gas \cite{condensation, vortex,
imbalance,Collective_Excitation,Duke}. At sufficiently low
temperatures, fermionic atoms are known to form pairs under an
attractive interaction. The interaction strength can be manipulated
by tuning magnetic fields near Feshbach resonance, which is
characterized by a detuning energy between the close channel bound
state energy and the open channel collision threshold. A positive
detuning leads to a negative scattering length, in this regime
paired atoms are loosely bound. Upon negative detuning, the
scattering length becomes positive and atoms can form bound
molecules, which could further condense into a BEC state.  Unlike
bosonic molecules, the statistics of interacting fermionic atoms is
dictated by Pauli exclusion principle, the ground state is thus made
up of a large number of modes, even at zero temperature. One usually
uses a BCS state to approximate the ground state at which fermions
are paired up according to their natural orbits \cite{deGennes}.
This is very different from BEC formed by pure bosons at zero
temperature, which is well described by a single mode wave function.
We may however expect, upon a strong enough interaction, paired
fermionic atoms become so tightly bound that they look just like
bosons \cite{Rink}. In that case, one natural question to ask is,
how alike are a fermionic pair and a boson? In this paper, we
address this question by constructing a Cooper pair creation
operator and examine its bosonic properties across resonance.

One fundamental feature that distinguishes fermions from bosons is
the commutation relation between their creation and annihilation
operators. For bosons, the commutator $\left[ C,C^\dag \right]$ is
one, while the anticommutator $\left\{ C,C^\dag \right\}$ is one for
fermions. For composite two-particles, the corresponding commutator
is not exactly one \cite{combescot,normalization,Law}. A useful
indicator measuring the deviation from the bosonic commutation
relation is the \textit{M}-pairs normalization factor $\chi_M$
defined by: $\left\langle 0 \right| C^M C^{\dag M}\left|
0\right\rangle=M!\chi_M$, where $\left| 0\right\rangle$ is the
vacuum state. The value of $\chi_M$ reflects the correction of Bose
enhancement factor, and was used to study ground state excitons
statistics \cite{combescot,normalization} and the connection to
quantum entanglement \cite{Law}. The key quantity was shown to be
the ratio $\chi_{M+1}/\chi_M$ which goes to one for a perfect boson.
This ratio will be one of our primary indicators of the bosonic
characters of Cooper pairs.

However, there has been an ambiguity in defining the explicit form
of a Cooper pair wave function. Ortiz \textit{et al.} \cite{Ortiz}
have given a discussion at length on this matter. In \cite{Yang1},
Yang showed that off-diagonal long range order exists in a
superconducting state, and is characterized by a dominant
eigenvector of the two-particle density matrix. The eigenvector is
sometimes recognized as a Cooper pair wave function. On the other
hand, the pair projection wave function of a BCS state
\cite{Randeria} is also a candidate. Both Cooper pair wave functions
will be examined in this paper. Their bosonic characters are
compared and we shall discuss their suitability as a bosonic mode in
a Fermi gas.

In this paper, we employ the one-channel approach to discuss the
crossover phenomena at zero temperature
\cite{Ortiz,Manini,analytic,Parish}. Specifically, a BCS state will
be used as our ground state wave function
\begin{eqnarray} \label{eq:BCS}
\left| \Phi \right\rangle=\prod_n \left( \tilde u_n+\tilde v_n
\alpha_n^\dag \beta_n^\dag \right) \left|0\right\rangle .
\end{eqnarray}
Here $\alpha_n$ and $\beta_n$ are the annihilation operators of two
spin components of fermonic atoms, $n$ denotes the quantum number of
pairing orbit, and $\tilde u_n$ and $\tilde v_n$ are amplitudes
subjected to normalization constraint $|\tilde u_n|^2+|\tilde
v_n|^2=1$. The number of atoms in each spin component is given by
$N=\sum\nolimits_n |\tilde v_n|^2$. In this paper, we will use the
solution of $\tilde u_n$, $\tilde v_n$ in homogeneous systems. For
trapped systems, the amplitudes can be determined by methods
described in Refs. \cite{German,Pong}.

\section{Bosonic tests}
We begin by reviewing some tests on the bosonic characters of a
particle operator. Consider the annihilation operator of a composite
particle defined by
\begin{eqnarray} \label{eq:COMPOSITE}
C=\sum\nolimits_n \lambda_n \alpha_n \beta_n ,
\end{eqnarray}
where $\sum_n |\lambda_n|^2=1$. The operator $C$, when sandwiched
by the BCS state given in (\ref{eq:BCS}), has the following
properties,
\begin{eqnarray}
\left\langle {C^\dag  C} \right\rangle  &=& \left|{\sum\nolimits_n
{\lambda_n \tilde u_n^* \tilde v_n } } \right|^2  +
\sum\nolimits_n {\left| {\lambda _n } \right|^2
 \left| {\tilde v_n } \right|} ^4 , \label{eq:FRACTION}\\
\left\langle {\left[ {C,C^\dag  } \right]} \right\rangle  &=&
\sum\nolimits_n \left|\lambda_n\right|^2\left(\left|\tilde u_n
\right|^2-\left|\tilde v_n \right| ^2 \right) \le 1 .
\label{eq:COMMUTATOR}
\end{eqnarray}
Eq. (\ref{eq:FRACTION}) gives the number of composite particles
existing in the gas. To quantify how `bosonic' the $C$ molecule is,
we study the commutator $\left[C,C^\dag\right]$. Note however only
expectation value of the commutator is given in Eq.
(\ref{eq:COMMUTATOR}), not the commutator itself. How close the
expectation value to unity is a necessary but not sufficient
condition for $C$ to be bosonic. To actually compare $C$ with a pure
boson operator, we adopt the views pointed out in
Refs.\cite{combescot,normalization,Law}. It was suggested that the
bosonic characters should be quantified by the normalization factor
$\chi_M$, where
\begin{eqnarray}
\left\langle 0 \right| C^MC^{\dag M} \left| 0\right\rangle=M!
\chi_M.
\end{eqnarray}
$\chi_M$ is obviously 1 if $C$ is a perfect boson. It is often more
convenient to look at the ratio $\chi_{M+1}/\chi_M$, since
\begin{eqnarray}
C\left| M
\right\rangle=\sqrt{\frac{\chi_M}{\chi_{M-1}}}\sqrt{M}\left|
M-1\right\rangle+\left| \epsilon_M \right\rangle ,
\label{eq:ENHANCEMENT}
\end{eqnarray}
where $|\epsilon_M\rangle$ is a correction term orthogonal to
$|M-1\rangle$, and it has the norm given by
\begin{eqnarray}
\left\langle \epsilon_M | \epsilon_M
\right\rangle=1-M\frac{\chi_M}{\chi_{M-1}}+(M-1)\frac{\chi_{M+1}}{\chi_M}.
\label{eq:EPSILON}
\end{eqnarray}
So we see that the ratio $\chi_{M+1}/\chi_M$ plays the role of a
correction of Bose enhancement factor with respect to a many body
state. It tells us how the gas differs from being bosonic, when one
more pair of atoms is added to or removed from a \textit{M}-pairs
gas.  The closer it is to one, the less a correction it is. The
criterion of a perfect boson is $\chi_{M+1}/\chi_M=1$,  only then is
$\langle \epsilon_M | \epsilon_M \rangle$ zero. In this paper we
will examine the case with $M=N$, which is the number of atoms of
one of the spin components in the gas.

In the case of fermions and if $N \gg 1$, Combescot \textit{et
al.} \cite{normalization} have shown that
\begin{eqnarray} \label{eq:FN+1}
\frac{\chi_{N+1}}{\chi_N} \approx \frac{N}{z_0},
\end{eqnarray}
where $z_0$ can be solved from the equation \cite{normalization}
\begin{eqnarray} \label{eq:SOLVE_Z0}
\frac{N}{z_0}=\sum\nolimits_n
\frac{\left|\lambda_n\right|^2}{1+z_0\left|\lambda_n\right|^2}.
\end{eqnarray}
From this last equation the ratio can be solved numerically. In
Refs. \cite{combescot,normalization}, the correction factor in
(\ref{eq:ENHANCEMENT}) has been studied in exciton systems. Here we
will apply Eq. (\ref{eq:SOLVE_Z0}) to atomic Cooper pairs with
$\lambda_n$ defined in the next section. One of the general features
is that the $\chi_{N+1}/\chi_N$ would deviate more from unity when
the density of atoms increases. This is because when the pair
density is large, atoms within each pair would see the Pauli effect
from atoms in nearby pairs.

It is useful to indicate the meaning of Eqs. (\ref{eq:FN+1}) and
(\ref{eq:SOLVE_Z0}) through a simple model. Let us consider
$|\lambda_k|^2=(\sum_{k'<k_C})^{-1}$ for $|k|<k_C$ and zero
otherwise, where $k_C$ is the extension of the wave function in
momentum space, such that the two particle wave function has a
spatial radius $\sim 1/k_C$. It can then be shown that Eq.
(\ref{eq:SOLVE_Z0}) reads
\begin{eqnarray} \label{eq:Z0_MEANING}
\frac{\chi_{N+1}}{\chi_N}\approx\frac{N}{z_0}=1-\frac{N}{V}\frac{6\pi^2}{k_C^3}.
\end{eqnarray}
Noting that $1/k_C^3$ is the spatial volume of our pair wave
function, $N$ is the maximum number of pairs in a total volume $V$,
the second term in Eq. (\ref{eq:Z0_MEANING}) is thus the volume
occupied by all Cooper pairs over the total volume. In the BCS limit
where a Cooper size is large ($k_C\sim$ Fermi momentum $k_F$,
$\frac{N}{V}\approx k_F^3/6\pi^2$), paired atoms are Pauli blocked
by atoms in between, preventing a bosonization, and hence $N/z_0$ is
nearly zero. While in the BEC limit where $k_C \rightarrow \infty$,
each Cooper pair is essentially isolated from each other, and this
gives $N/z_0 \to 1$ \cite{normalization}.

\section{Cooper pair wave functions}
\label{sec:CHOOSE_LAMBDA} We now discuss two choices of $\lambda_n$
in defining the Cooper pair wave functions. First, it was shown in
\cite{Yang1} that the long range correlation
($|\textbf{r}_1-\textbf{r}_1'|\rightarrow \infty$,
$|\textbf{r}_2-\textbf{r}_2'|\rightarrow \infty$ ) in a paired state
is reflected in the eigenvalue decomposition of the two particle
density matrix
\begin{eqnarray} \label{eq:EXPANSION}
G (\textbf{r}_1',\textbf{r}_2';\textbf{r}_1,\textbf{r}_2)
= \left\langle \psi_\alpha^\dag \left( \textbf{r}_1'\right)
\psi_\beta^\dag \left( \textbf{r}_2'\right)\psi_\beta
\left( \textbf{r}_2\right)\psi_\alpha \left( \textbf{r}_1\right)
\right\rangle \;\;\;\;\;\; \nonumber \\
\approx \left(\sum_m \left|\tilde u^*_m \tilde v_m
\right|^2\right) \tilde
\phi(\textbf{r}_1,\textbf{r}_2)\tilde\phi^*(\textbf{r}_1',
\textbf{r}_2'),\;\;\;\;\;\;
\end{eqnarray}
where $\psi_\alpha\left(\textbf{r}\right)=\sum\nolimits_n
f_n\left(\textbf{r}\right)\alpha_n$ and
$\psi_\beta\left(\textbf{r}\right)=\sum\nolimits_n
f^*_n\left(\textbf{r}\right)\beta_n$ are the field operators of
the respective species, and $\{f_n\}$ is an orthonormal set of
natural orbits. The mode function $\tilde \phi$ can be written as
\begin{eqnarray} \label{eq:FUNCTION_UV}
\tilde \phi(\textbf{r}_1,\textbf{r}_2)&=& \frac{\sum\nolimits_{n}
\tilde u_n^* \tilde v_n
f_n^*(\textbf{r}_2)f_n(\textbf{r}_1)}{\sqrt{\sum_m \left|\tilde
u^*_m \tilde v_m \right|^2}}
\end{eqnarray}
in terms of the natural orbits $\{f_n\}$. $\tilde
\phi(\textbf{r}_1,\textbf{r}_2)$ is often considered as a Cooper
pair wave function and $2\sum\nolimits_n \left|\tilde u^*_n
\tilde v_n \right|^2$ is the number of atoms that condense into
Cooper pairs. We shall thus construct $C$ with respect to this
wave function. $\tilde \phi(\textbf{r}_1,\textbf{r}_2)$ is
obviously associated with $C$ through
\begin{eqnarray} \label{eq:LAMBDA_UV}
\lambda_n=\frac{\tilde u_n \tilde v_n^*}{\sqrt{\sum\nolimits_n
|\tilde u_n \tilde v_n^*|^2 }},
\end{eqnarray}
which gives the explicit expressions
\begin{eqnarray}
\left\langle {C^\dag  C} \right\rangle  &=& \sum_n |
\tilde u_n^* \tilde v_n |^2 + \frac{\sum\nolimits_n \left|
\tilde u_n \tilde v_n^* \right|^2 \left|\tilde v_n \right|^4}
{\sum_n | \tilde u_n^* \tilde v_n |^2}  \label{eq:FRACTION_UV},\\
 \left\langle \left[ C,C^\dag \right] \right\rangle  &=&
 \frac{\sum\nolimits_n \left|\tilde u_n \tilde v_n^* \right|^2
 \left(\left|\tilde u_n \right|^2-\left|\tilde v_n \right| ^2 \right)}
 {\sum_n |\tilde u_n^* \tilde v_n |^2}. \label{eq:COMMUTATOR_UV}
\end{eqnarray}
The second term in $\left\langle {C^\dag  C} \right\rangle$ is
smaller than unity, while the first term is of order $N$. So as long
as the number of particles is large, the second term can be dropped.
In the large $N$ limit,  $\left\langle {C^\dag  C} \right\rangle$ is
just the eigenvalue given in (\ref{eq:EXPANSION}). We remark that
Eq. (\ref{eq:LAMBDA_UV}) was also recognized by Leggett
\cite{Leggett} as a form of Cooper pair wave function, and recently
Salasnich {\it et al.} have made use of the definition to calculate
the condensate fraction \cite{Manini}.

There is however another way of defining a Cooper pair based on the
single pair projection from a BCS state \cite{Rink, Randeria,
Ortiz,ho}. By expressing the BCS state as
\begin{eqnarray} \label{eq:COHERENT_EXPANSION}
\left| \Phi \right\rangle = \left( \prod_n \tilde u_n
\right)\sum_{j=0}^\infty \left( \sum_n \left|\tilde v_n/\tilde
u_n \right|^2\right)^{\frac{j}{2}}\frac{C'^{\dag
j}}{j!}\left|0\right\rangle,
\end{eqnarray}
$C'^\dag$ would then be a Cooper pair creation operator. It can be
shown that $C'$ takes the form \cite{Rink, Randeria, Ortiz,ho}:
\begin{eqnarray}
C' &=& \sum\nolimits_n \lambda'_n \alpha_n \beta_n, \\
\lambda'_n &=&\frac{\tilde v_n/\tilde u_n}{\sqrt{\sum\nolimits_m
|\tilde v_m/\tilde u_m|^2}}. \label{eq:LAMBDA_V/U}
\end{eqnarray}
Applying the previous procedures on $C'$, we have
\begin{eqnarray}
\left\langle C'^\dag  C' \right\rangle &=& 1+\frac{\left\langle N
\right\rangle^2 -2\left\langle N \right\rangle+ \sum\nolimits_n |
\tilde u_n^* \tilde v_n |^2}{\sum\nolimits_n \left|
\tilde v_n/\tilde u_n\right|^2 } ,\label{eq:FRACTION_V/U} \\
\left\langle \left[ C',C'^\dag \right] \right\rangle  &=&
-1+2\left\langle N\right\rangle/\sum\nolimits_n
\left|\frac{\tilde v_n}{\tilde u_n}\right|^2.
\label{eq:COMMUTATOR_V/U}
\end{eqnarray}
Note that the BCS state given in Eq. (\ref{eq:COHERENT_EXPANSION})
is in fact a coherent state if $C'$ is perfectly bosonic.

\section{Results in a uniform gas}
Before proceeding, let's recap some familiar results in a
homogeneous BCS gas. The natural orbits are the plane wave mode
$f_\textbf{k}\left(\textbf{r}\right)=e^{i\textbf{k}\cdot
\textbf{r}}/\sqrt{V}$ and the occupation amplitudes are given by
\begin{eqnarray} \label{eq:UV}
\left( {\begin{array}{*{20}c}
   {\tilde u_k }  \\
   {\tilde v_k }  \\
\end{array}} \right)= \frac{1}{\sqrt 2}\sqrt{ 1 \pm \frac{k^2-2\mu}
{\sqrt {(k^2 - 2\mu )^2 + 4\Delta ^2 }} }
\end{eqnarray}
where $\Delta=4\pi a\left \langle\psi_\beta (\textbf{r}) \psi_\alpha
(\textbf{r}) \right\rangle=-4\pi a \sum_\textbf{k} \tilde u_k \tilde
v_k$ is the pairing gap, $\mu$ is the chemical potential, and $a$ is
the scattering length. $\mu$, $\Delta$, $a$ and the atom density
$\rho$ of each species are related by a regularized gap equation and
a number equation,
\begin{eqnarray}
-\frac{1}{4\pi a} &=&\int \frac{d^3k}{(2\pi)^3}\left(
\frac{1}{2\sqrt{(k^2/2-\mu)^2+\Delta^2}}-\frac{1}{k^2}\right)
\label{eq:GAP_EQUATION} \\
\rho&=&\frac{N}{V} = \int \frac{d^3k}{\left(2\pi\right)^3}
\left| \tilde v_k \right|^2 \label{eq:NUMBER_EQUATION}
\end{eqnarray}
where the integration can be expressed in terms of elliptic
integrals \cite{analytic}. The Fermi momentum is defined as
$k_F=\left(6\pi^2\rho \right)^{1/3}$, which is the reciprocal of the
interatomic distance. An important dimensionless parameter is
$\left( k_Fa\right)^{-1}$. The BCS limit is indicated by $\left(
k_Fa\right)^{-1} \ll -1$, the BEC limit corresponds to $\left(
k_Fa\right)^{-1} \gg 1$, and the crossover occurs at $\left(
k_Fa\right)^{-1}=0$ \cite{analytic,Manini}. Some integrals used are
listed in the Appendix for reference.

\begin{figure}
\includegraphics [width=5.5 cm] {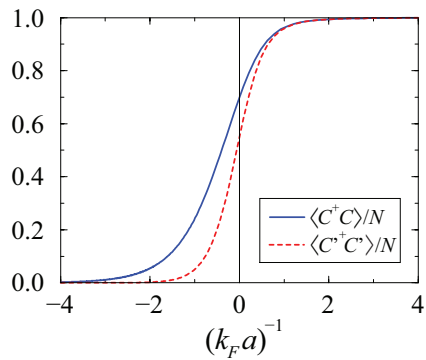}
\caption{\label{fig:FRACTION} (Color online) The fraction of
condensate particles $\langle C^\dag C\rangle/N$ with two different
definitions: (a) $\lambda_k \sim \tilde u_k \tilde v_k$ (solid line)
(Eq.(\ref{eq:FRACTION_UV})) and (b)$\lambda_k' \sim \tilde
v_k/\tilde u_k$ (red dashed line) (Eq.(\ref{eq:FRACTION_V/U})).}
\end{figure}

Using Eq. (\ref{eq:UV}) for $\tilde u_k$ and $\tilde v_k$, we
evaluate Eq. (\ref{eq:FRACTION_UV}, \ref{eq:COMMUTATOR_UV}) and Eq.
(\ref{eq:FRACTION_V/U}, \ref{eq:COMMUTATOR_V/U}). In Fig.
\ref{fig:FRACTION} we plot the fraction of condensation in the gas,
as a function of the dimensionless parameter $(k_Fa)^{-1}$. With
either choice of $\lambda_k$, the fraction goes to one in the BEC
limit $(k_Fa)^{-1} \gg 1$. Notice that $\left\langle C^\dag
C\right\rangle/N$ is an appreciably higher fraction than
$\left\langle C'^\dag C'\right\rangle/N$, showing a dominant
condensation of atoms into the pair wave function defined in
(\ref{eq:FUNCTION_UV}).

\begin{figure}
\includegraphics [width=5.5 cm] {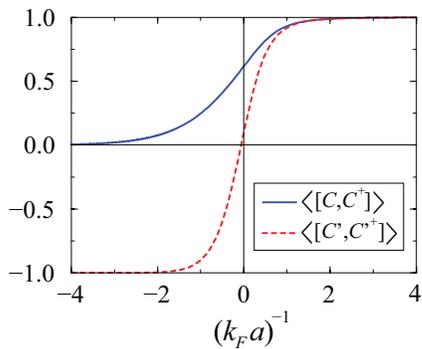}
\caption{\label{fig:COMMUTATOR} (Color online) Expectation value of
the commutator $\langle \left[C,C^\dag \right]\rangle$ with two
different definitions: (a) $\lambda_k \sim \tilde u_k \tilde v_k$
(solid line) (Eq. (\ref{eq:COMMUTATOR_UV})) and (b)$\lambda_k' \sim
\tilde v_k/\tilde u_k$ (red dashed line) Eq.
(\ref{eq:COMMUTATOR_V/U})).}
\end{figure}

The expectation value of the commutator $\langle \left[
C,C^\dag\right] \rangle$ as a function of $(k_Fa)^{-1}$ is shown in
Fig. \ref{fig:COMMUTATOR}. Again both definitions of $\lambda_k$
give unity in the BEC limit, but Eq. (\ref{eq:COMMUTATOR_UV}) is
always closer to one than Eq. (\ref{eq:COMMUTATOR_V/U}).

\begin{figure}
\includegraphics [width=5.5 cm] {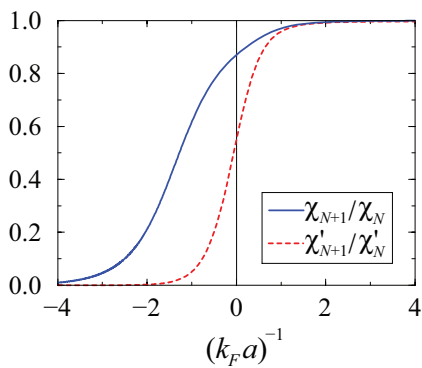}
\caption{\label{fig:FN} (Color online) The quantity
$\chi_{N+1}/\chi_N$ with two different definitions: (a) $\lambda_k
\sim \tilde u_k \tilde v_k$ (blue solid line) and (b)$\lambda_k'
\sim \tilde v_k/\tilde u_k$ (red dashed line).}
\end{figure}

Next we calculate the factor $\chi_{N+1}/\chi_N$. By solving Eq.
(\ref{eq:SOLVE_Z0}) numerically for $\lambda_k$ and $\lambda_k'$, we
obtain the ratios $\chi_{N+1}/\chi_N$ and $\chi_{N+1}'/\chi_N'$ from
Eq. (\ref{eq:FN+1}). These ratios are shown in Fig. \ref{fig:FN} as
a function of $(k_F a)^{-1}$. We see that $\chi_{N+1}/\chi_N$ is
closer to one throughout the transition region. Together with the
tests based on expectation values above, $\lambda_k$ defined Eq.
(\ref{eq:LAMBDA_UV}) seems to be a more suitable choice for the
bosonic description of paired atoms.

Our calculations indicate an interesting region roughly at $-2
\lesssim (k_Fa)^{-1} \lesssim 2$ where Cooper pairs transit from
being non-bosonic to bosonic. Note that it does not require $(k_F
a)^{-1} \gg 1$ for the emergence of bosonic character. At $(k_F
a)^{-1} =1$, the fraction of condensation $\langle C^\dag C \rangle
/ N$ is already 95\%, $\langle \left[C,C^\dag\right]\rangle\sim
0.94$ and $\chi_{N+1}/\chi_N \sim 0.97$. In particular at the point
where the chemical potential $\mu=0$ ($(k_F a)^{-1}\approx0.553$)
\cite{analytic}, which is sometimes recognized as the boundary
between BEC and BCS regimes \cite{Leggett2,Chen}, we have $\langle
\left[ C,C^\dag \right] \rangle=0.835$, $\chi_{N+1}/\chi_N=0.937$.
The use of definition (\ref {eq:LAMBDA_V/U}) gives slightly weaker
numbers, but a narrower transition region.

\begin{figure}
\includegraphics [width=4.5 cm] {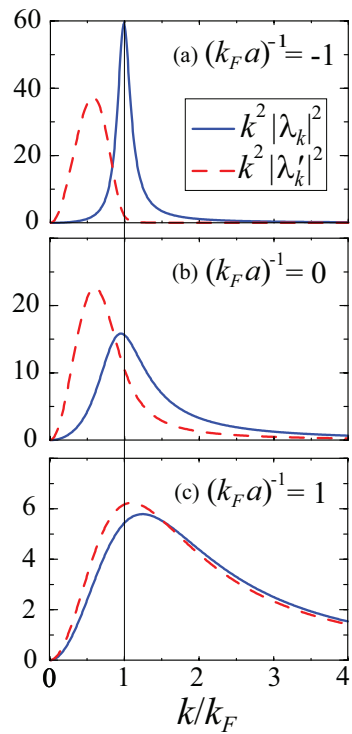}
\caption{\label{fig:fig4} (Color online) Momentum distribution (in
units of $k_F$) of two different Cooper pair wave functions,
$\lambda_k \sim \tilde u_k \tilde v_k$ (blue line) and
$\lambda_k'\sim \tilde v_k / \tilde u_k$ (red dashed line). In (a)
the BCS limit, two definitions differ significantly, while in (c)
the BEC limit, two definitions overlap (see text).}
\end{figure}

\section{Conclusion}

To conclude, three indicators were used to quantify the bosonic
characters of a Cooper pair in an interacting Fermi gas : (a)
$\langle C^\dag C \rangle$, (b) $\langle \left[C, C^\dag \right]
\rangle$, (c) $\chi_{N+1}/\chi_N$. Two different definitions of a
Cooper pair were examined, $\lambda_n'\propto \tilde v_n/\tilde u_n$
and $\lambda_n\propto \tilde u^*_n\tilde v_n$. Our calculations
suggest that the latter one provides a better description of the
Cooper pairs as bosonic particles. Moreover, as the fraction of
composite particles goes to one in the BEC limit, the gas is
basically in its simplest single mode. It appears that using either
one of the two definitions makes little differences in the strong
coupling regime $(k_F a)^{-1} \geq 1$. This is consistent with the
results in Ref. \cite{Ortiz}, in which the authors addressed the
similarity of both definitions. As shown in Fig. 4, the difference
between ({\ref{eq:LAMBDA_UV}) and (\ref{eq:LAMBDA_V/U}) on the BCS
side is that, the former only takes into account of a few momentum
states \textit{on} the Fermi surface, whereas the latter one takes
the average of all states \textit{inside} the Fermi sphere. In a
weakly interacting gas, only atoms lying on the Fermi surface
interact effectively, the composite particle based on
({\ref{eq:LAMBDA_UV}) would thus be more bosonic since it takes into
account the dominant correlated states. In the deep BEC limit, the
Fermi sphere is totally smeared out, and either choice of
$\lambda_n$ would weigh different momentum states on an almost equal
footage, resulting in the merge of two different pictures.

If the system is not in the BEC limit, we have shown that the
bosonic character of a Cooper pair depends on how the pair wave
function is defined. Our work here is an attempt to identify what
definition is more effective to reveal a Cooper pair as a boson.
From an experimental point of view, it is interesting to search for
observables associated with $C$ or $C'$, so that one can probe the
the quantum statistics of Cooper pairs directly. We also remark that
our method can in principle be extended to nonuniform systems.
However, the coefficients $\tilde u_n$ and $\tilde v_n$, which are
calculated from the natural orbits \cite{German,Pong}, do not have
closed forms for analytical discussions. The question how a trapping
potential affects the bosonic properties of Cooper pairs is an open
problem for future studies.

\section*{Acknowledgement}
This work is supported in part by the Research Grants Council of the
Hong Kong Special Administrative Region, China (Project No. 401406).

\section*{Appendix}
We list here some integrals used in this paper. A detail analysis
can be found in the paper by Marini \textit{et al.}
\cite{analytic} and the paper by Ortiz \textit{et al.}
\cite{Ortiz}. In the following list, we adopt the following change of variables,
\begin{eqnarray}
x_0=\mu/\Delta , \;\;\; x^2=k^2/(2\Delta)
\end{eqnarray}
and introduce shorthand notations
\begin{eqnarray}
\xi_x&=&x^2-x_0 ,\;\;\; E_x=\sqrt{\xi_x^2+1},\\
\kappa^2&=&\frac{1}{2}\left( 1+x_0/\sqrt{x_0^2+1}\right) ,\\
q&=&-x_0/\sqrt{x_0^2+1}.
\end{eqnarray}
So we have $\int k^2 dk = \left(2\Delta\right)^{3/2} \int x^2dx$.
Some integrals that appeared in our calculation are listed below
\begin{eqnarray}
-\frac{1}{4\pi a}&=&\frac{\sqrt{2\Delta}}{2\pi^2}\int_0^\infty x^2dx
\left(\frac{1}{E_x}-\frac{1}{x^2} \right) \\
\frac{N}{V}&=&\frac{\left(2\Delta\right)^{3/2}}{4\pi^2 }
\int_0^\infty x^2dx \left(1-\frac{\xi_x}{E_x} \right) \\
\int \frac{d^3k}{(2\pi)^3}\left|\tilde u_k \tilde v_k\right|^2
&=&\frac{(2\Delta)^{3/2}}{8\pi^2}\int_0^\infty \frac{x^2dx}{E_x^2} \\
\int \frac{d^3k}{(2\pi)^3}\left|\frac{\tilde v_k}{ \tilde u_k}\right|^2
&=& \frac{(2\Delta)^{3/2}}{2\pi^2}\int_0^\infty x^2 dx \frac{E_x-\xi_x}{E_x+\xi_x}
\end{eqnarray}
The integrals are expressed in terms of $K$ and $E$, which are
respectively the complete elliptic integral of the first and second
kind. We also need $P_n$, Legendre function of the first kind of
degree $n$.
\begin{eqnarray}
&& \int_0^\infty x^2dx \left( \frac{1}{E_x}-\frac{1}{x^2}\right)
=\frac{K(\kappa^2)-2E(\kappa^2)}{\left( x_0^2+1\right)^{-1/4}}\\
&&\int_0^\infty \frac{x^2dx}{E_x^2}
=\frac{\pi}{2\sqrt{2}}\sqrt{x_0+\sqrt{x_0^2+1}} .
\end{eqnarray}
\begin{eqnarray}
&& \int_0^\infty x^2 dx\left(1-\frac{\xi_x}{E_x}\right)
%&& = \frac{\left(\sqrt{x_0^2+1}-x_0
%\right)K(\kappa^2)+2x_0E(\kappa^2)}
%{3\left(x_0^2+1\right)^{-1/4}} \nonumber \\
= \frac{(1+q)K(\kappa^2)-2qE(\kappa^2)}{3(x_0^2+1)^{-3/4}} \\
&& \int_0^\infty x^2 dx \frac{E_x-\xi_x}{E_x+\xi_x}  \nonumber \\
&& =\frac{2\pi}{35}\frac{\left(q^2-5\right) P_{3/2}\left( q\right)
+4qP_{1/2}\left(q\right)}{(x_0^2+1)^{-7/4}}
\end{eqnarray}
Choosing $\lambda_k=\tilde u_k \tilde v_k /\sqrt{\sum_{\textbf{k}'} |\tilde u_{k'} \tilde v_{k'}|^2}$, Eq.
(\ref{eq:SOLVE_Z0}) is solved with the help of the following integral,
\begin{eqnarray}
\int_0^\infty \frac{x^2dx}{bE_x^2+1} &=&
\frac{\pi\sqrt{x_0+\sqrt{x_0^2+1+1/b}}}{\sqrt{8}\sqrt{b(b+1)}} ,
\end{eqnarray}
where $b$ is a positive real number. For $\lambda_k'=(\tilde
v_k/\tilde u_k) /\sqrt{\sum_{\textbf{k}'} |\tilde v_{k'}/\tilde u_{k'}|^2}$, Eq.
(\ref{eq:SOLVE_Z0}) is numerically integrated and solved.


\begin{references}

\bibitem{condensation} C. A. Regal, M. Greiner,
and D. S. Jin, Phys. Rev. Lett. \textbf{92}, 040403 (2004).
\bibitem{vortex} M. W. Zwierlein, C. A. Stan, C. H. Schunck,
S. M. F. Raupach, A. J. Kerman, and W. Ketterle, Phys. Rev. Lett.
{\bf 92}, 120403 (2004); M. W. Zwierlein, J. R. Abo-Shaeer, A.
Schirotzek, C. H. Schunck, and W. Ketterle, Nature \textbf{435},
1047 (2005).
\bibitem{imbalance} Martin W. Zwierlein, Andre Schirotzek, Christian H. Schunck,
and Wolfgang Ketterle, Science \textbf{311}, 492 (2006).
\bibitem{Collective_Excitation} M. Bartenstein, A. Altmeyer, S. Riedl, S. Jochim, C. Chin, J. Hecker Denschlag, R. Grimm, Phys. Rev. Lett. \textbf{92}, 203201 (2004).
\bibitem{Duke} K. M. O'Hara, S. L. Hemmer, M. E. Gehm, S. R. Granade, J. E. Thomas, Science \textbf{298}, 2179 (2002).
\bibitem{deGennes} P. G. de Gennes, \textit{Superconductivity of Metals and Alloys}
(W.A. Benjamin INC.,1966); J. B. Ketterson and S. N. Song, \textit{Superconductivity}
(New York, Cambridge University Press, 1999).
\bibitem{Rink} P. Nozi\'eres, S. Schmitt-Rink, J. Low  Temp. Phys. \textbf{59}, 195 (1985).
\bibitem{combescot} M. Combescot and C. Tanguy, Europhys. Lett. {\bf 55}, 390 (2001).
\bibitem{normalization} M. Combescot, X. Leyronas, and C. Tanguy,
Eur. Phys. J. B \textbf{31}, 17 (2003).
\bibitem{Law} C. K. Law, Phys. Rev. A {\bf 71}, 034306 (2005).
\bibitem{Ortiz} G. Ortiz and J. Dukelsky, Phys. Rev. A \textbf{72}, 043611 (2005);
G. Ortiz and J. Dukelsky, cond-mat/0604236.
\bibitem{Yang1} C. N. Yang, Rev. Mod. Phys. \textbf{34}, 694 (1962).
\bibitem{Randeria} M. Randeria, in \textit{Bose Einstein Condensation},
edited by A. Griffin, D. W. Snoke and S. Stringari (Cambridge
University Press 1995).
\bibitem{ho} Roberto B. Diener and Tin-Lun Ho, cond-mat/0404517.
\bibitem{Manini} Luca Salasnich, Nicola Manini, Alberto Parola, Phys. Rev. A \textbf{72}, 023621 (2005).
\bibitem{analytic} M. Marini, F. Pistolesi, and G.C. Strinati, Eur. Phys. J. B \textbf{1}, 151 (1998).
\bibitem{Parish}Meera M. Parish, Bogdan Mihaila, Eddy M. Timmermans,
Krastan B. Blagoev, and Peter B. Littlewood, Phys. Rev. B {\bf 71},
064513 (2005).
\bibitem{German} P.-G. Reinhard, M. Bender, K. Rutz, and J. A. Maruhn,
Z. Phys. \textbf{A 358}, 277 (1997).
\bibitem{Pong} Y.H. Pong and C. K. Law, Phys. Rev. A \textbf{74} 013618 (2006).
\bibitem{Leggett} A. J. Leggett, J. de Physique \textbf{41}, C7-19 (1980).
\bibitem{Leggett2} A. J. Leggett, in {\it Modern Trends in the Theory of Condensed Matter},
edited by A. Pekalski and R. Przystawa (Springer-Verlag, Berlin,
1980).
\bibitem{Chen} Qijin Chen, Jelena Stajic, Shina Tan, and K. Levin,
Phys. Rep. \textbf{412}, 1 (2005).

\end{references}
\end{document}